# Nonlinear ac stationary response and dynamic magnetic hysteresis of quantum uniaxial superparamagnets

Yuri P. Kalmykov,[a] Serguey V. Titov,[b] and William T. Coffey[c]

[a] *Laboratoire de Mathématiques et Physique (LAMPS, EA 4217), Université de Perpignan Via Domitia, 52, Avenue de Paul Alduy, 66860 Perpignan Cedex, France*

[b] *Kotel'nikov Institute of Radioengineering and Electronics of the Russian Academy of Sciences, Vvedenskii Square 1, Fryazino, 141190, Russian Federation*

[c] *Department of Electronic and Electrical Engineering, Trinity College, Dublin 2, Ireland*

**ABSTRACT**

The nonlinear ac stationary response of uniaxial paramagnets and superparamagnets – nanoscale solids or clasters with spin number $S \sim 10^0$ - $10^4$ – in superimposed uniform ac and dc bias magnetic fields of arbitrary strength, each applied along the easy axis of magnetization, is determined by solving the evolution equation for the reduced density matrix represented as a *finite* set of three-term differential-recurrence relations for its *diagonal* matrix elements. The various harmonic components of the magnetization, dynamic magnetic hysteresis loops, etc. are then evaluated via matrix continued fractions indicating a pronounced dependence of the nonlinear response on $S$ arising from the quantum spin dynamics. In the linear response approximation, the results concur with existing solutions.

# I. INTRODUCTION

Nanomagnetism is a rapidly expanding area of research with many novel applications particularly in information storage [1] and in medicine, e.g., in hyperthermia occasioned by induction heating of nanoparticles [2,3]. Here single domain ferromagnetic particles exhibit essentially classical behavior while smaller entities such as free nanoclusters made of many atoms, molecular clusters, and single molecule magnets exhibit pronounced quantum effects. Now, due to their large magnetic dipole moment (~10–$10^5$ Bohr magneton $\mu_B$), the magnetization relaxation of nanomagnets driven by an ac field will exhibit a pronounced field and frequency dependence which is significant in diverse physical applications. These include nonlinear dynamic susceptibilities [4-7], stochastic resonance [8], the dynamic magnetic hysteresis [9-12], etc. In general, however, the nonlinear response to an external field invariably poses a difficult problem because that response will always depend on the *precise* nature of the stimulus. Thus, no *unique* response function valid for all stimuli exists unlike in the linear response to a weak magnetic field. These difficulties are compounded in quantum spin systems such as molecular magnets and nanoclusters, where both the field and frequency dependence of the dynamic response to an ac driving field (which is our main concern here) differ profoundly from their classical counterparts due to tunneling effects [4].

In the context of linear response theory, spin relaxation of nanomagnets with arbitrary spin number *S* was usually treated via the evolution equation for the spin density matrix using the second order of perturbation theory in the spin bath coupling (see, e.g., [13-17]). In particular, Garanin [13] and García-Palacios *et al.* [16] gave a concise treatment of the longitudinal spin relaxation of *uniaxial* nanomagnets by proceeding from the quantum Hubbard operator representation of the evolution equation for the spin density matrix. This problem has also been treated [18-20] via the master equation for the distribution function of spin orientations in the representation (phase) space of the polar and azimuthal angles that is completely analogous [21-24] to the treatment of the magnetization relaxation of nanomagnets with classical superparamagnetic behavior via Fokker-Planck equation governing the evolution of the distribution function of magnetization orientations [25]. An important result of all these studies is that one can now accurately evaluate quantum effects in the linear dynamic susceptibility, reversal time of the magnetization, etc. of nanomagnets [16,18-20]. Furthermore, one can estimate the range of spin numbers *S*, where the crossover to classical superparamagnetic behavior of nanomagnets pertaining to a giant classical spin and that corresponding to the classical limit, $S \to \infty$, takes place (typically, this appears in the range $S \sim 20$-$50$ [14,17,19]). However, the results obtained in Refs. [13-20] using linear response theory cannot be applied to *nonlinear* phenomena such as the magnetization reversal in nanomagnets driven by a strong ac external magnetic field, nonlinear stochastic resonance, dynamic magnetic hysteresis (DMH), etc. because they are related to the *nonlinear* ac stationary response in the presence of thermal agitation. Hitherto, that response for quantum nanomagnets has been determined via perturbation theory (e.g., Ref. 4) by supposing that the external ac field is weak. In evaluating the nonlinear response of a nanomagnet to an ac field of *arbitrary strength*, perturbation theory is now no longer applicable. However, as we shall demonstrate, quantum effects in the nonlinear response of nanomagnets can be treated by generalizing methods developed for classical spins [26] (see also [25], Chap. 9).

As a generic model, we shall consider a uniaxial nanomagnet with arbitrary spin number $S$ subjected to superimposed spatially *uniform* dc and ac fields $\mathbf{H}_0$ and $\mathbf{H}(t) = \mathbf{H}\cos\omega t$, respectively, applied along the Z-axis, which is the easy axis of magnetization. Thus, the time-dependent Hamiltonian $\hat{H}_S(t)$ has the axially symmetric form

$$\beta\hat{H}_S(t) = -\frac{\sigma}{S^2}\hat{S}_Z^2 - \frac{\xi_0 + \xi\cos\omega t}{S}\hat{S}_Z, \qquad (1)$$

where $\hat{S}_Z$ is the operator associated with the Z-component of the spin [24], $\sigma$ is the dimensionless anisotropy constant, $\xi_0 = \beta S\hbar\gamma H_0$ and $\xi = \beta S\hbar\gamma H$ are the dc bias and ac field parameters, respectively, $\gamma$ is the gyromagnetic ratio, $\hbar$ is Planck's constant, and $\beta = (kT)^{-1}$ is the inverse thermal energy. This Hamiltonian comprises a uniaxial anisotropy term $-\sigma\hat{S}_Z^2/S^2$ plus the Zeeman term $-(\xi_0 + \xi\cos\omega t)\hat{S}_Z/S$. In particular, it represents an archetypal model for spin relaxation phenomena in molecular magnets, nanoclusters, etc. For large $S$, the Hamiltonian Eq. (1) describes the magnetization relaxation of classical superparamagnets like magnetic nanoparticles [16]. Moreover, the time-independent Hamiltonian $\beta\hat{H}_S = -\sigma\hat{S}_Z^2/S^2 - \xi_0\hat{S}_Z/S$ is commonly used, e.g., to describe the magnetic properties of $Mn_{12}$ clusters with $S = 10$, $\sigma T/S^2 = 0.6 \div 0.7\,\text{K}$ [13,16]. In the standard basis of spin functions $|S,m\rangle$, which describe the states with definite spin $S$ and spin projection $m$ onto the Z-axis, i.e., $\hat{S}_Z|S,m\rangle = m|S,m\rangle$, this Hamiltonian has an energy spectrum with a double-well structure and two minima at $m = \pm S$ separated by a potential barrier. Notice that in strong bias fields, $\xi_0 > \sigma(2S-1)/S$, the barrier disappears. Now generally speaking, spin reversal can take place either by thermal activation or by tunneling or a combination of both. The tunneling may occur from one side of the barrier to the other between resonant, equal-energy states coupled by transverse fields or high-order anisotropy terms [13,16]. The evolution equation for the reduced density matrix $\hat{\rho}$ describing the spin relaxation of a uniaxial nanomagnet with the Hamiltonian $\hat{H}_S(t)$, Eq. (1), coupled to a thermal bath is

$$\frac{\partial \hat{\rho}(t)}{\partial t} + \frac{i}{\hbar}\left[\hat{H}_S(t), \hat{\rho}(t)\right] = \mathrm{St}\{\hat{\rho}(t)\}, \qquad (2)$$

In Eq. (2), the collision kernel operator $\mathrm{St}\{\hat{\rho}(t)\}$ characterizing the spin-bath interaction we will employ is given by (see Appendix A)

$$\mathrm{St}\{\hat{\rho}(t)\} = \sum_{\mu=-1}^{1} (-1)^\mu D_\mu \left( \left[ \hat{S}_\mu, \hat{\rho}(t) e^{\beta \hat{H}_S(t)/2} \hat{S}_{-\mu} e^{-\beta \hat{H}_S(t)/2} \right] \right. \\
\left. + \left[ e^{-\beta \hat{H}_S(t)/2} \hat{S}_{-\mu} e^{\beta \hat{H}_S(t)/2} \hat{\rho}(t), \hat{S}_\mu \right] \right). \qquad (3)$$

Here the square brackets denote the commutators, viz., $\left[\hat{A}, \hat{B}\right] = \hat{A}\hat{B} - \hat{B}\hat{A}$, $D_\mu$ are "diffusion" coefficients, $\hat{S}_0 = \hat{S}_Z$, $\hat{S}_{\pm 1} = \mp\left(\hat{S}_X \pm i\hat{S}_Y\right)/\sqrt{2}$ and $\hat{S}_X$, $\hat{S}_Y$, $\hat{S}_Z$ are, respectively, the spherical and Cartesian components of the spin [27]. The above model was proposed by Hubbard [28] by generalizing Redfield's derivation [29] of the density matrix evolution equation to time-dependent Hamiltonians $\hat{H}_S(t)$ (the original Redfield derivation [29] was limited to time-independent Hamiltonians $\hat{H}_S$). As shown in Appendix A, the Hubbard model [28] of the collision kernel $\mathrm{St}\{\hat{\rho}(t)\}$ in the *short bath correlation time approximation*, can be simplified to yield Eq. (3) [22,30]. This simplification implies that the correlation time $\tau_c$ characterizing the thermal bath is short enough to approximate the stochastic process originating in the bath by a Markov process thus qualitatively describing the spin relaxation in nanomagnets at least in the high temperature limit. In the parameter range, where the above approximation fails, e.g., throughout the very low temperature region, more general forms of the density matrix evolution equation must be used, e.g., those suggested in Refs. [13,14,16,17]. Using the above model, we will now calculate the nonlinear ac stationary response of a quantum uniaxial nanomagnet with arbitrary *S*. Furthermore, we will show that our results in the weak ac field approximation, $\xi \ll 1$, coincide with existing linear response solutions for quantum nanomagnets [16,18,19] while in the classical limit, $S \to \infty$, they correspond with those of Ref. [26] for classical spin systems.

## II. SOLUTION OF THE EVOLUTION EQUATION

For a uniaxial nanomagnet with the Hamiltonian Eq. (1), the evolution equation (2) for reduced density matrix becomes

$$\frac{\partial \hat{\rho}}{\partial t} = \frac{i}{\hbar \beta} \left\{ \frac{\sigma}{S^2}\left[\hat{S}_0^2, \hat{\rho}\right] + \frac{\xi_0 + \xi \cos \omega t}{S}\left[\hat{S}_0, \hat{\rho}\right] \right\} \\
+ D_\parallel \left( \left[\hat{S}_0, \hat{\rho}\hat{S}_0\right] + \left[\hat{S}_0 \hat{\rho}, \hat{S}_0\right] \right) \qquad (4) \\
- 2D_\perp \left\{ e^{-\frac{\sigma}{2S^2} - \frac{\xi_0 + \xi \cos \omega t}{2S}} \left[\hat{S}_{-1} e^{-\frac{\sigma}{S^2}\hat{S}_0} \hat{\rho}, \hat{S}_{+1}\right] + e^{\frac{\sigma}{2S^2} + \frac{\xi_0 + \xi \cos \omega t}{2S}} \left[\hat{S}_{+1} e^{\frac{\sigma}{S^2}\hat{S}_0} \hat{\rho}, \hat{S}_{-1}\right] \right\},$$

where we have introduced the notation $2D_\perp = D_{+1} = D_{-1}$ and $D_\parallel = D_0$ for the diffusion coefficients and have used the operator relations

$$e^{\frac{\sigma}{2S^2}\hat{S}_0^2 + \frac{\xi_0 + \xi\cos\omega t}{2S}\hat{S}_0} \hat{S}_{\pm 1} e^{-\frac{\sigma}{2S^2}\hat{S}_0^2 - \frac{\xi_0 + \xi\cos\omega t}{2S}\hat{S}_0}$$

$$= e^{-\frac{\sigma}{2S^2} \pm \frac{\xi_0 + \xi\cos\omega t}{2S}} e^{\pm\frac{\sigma}{S^2}\hat{S}_0} \hat{S}_{\pm 1},$$

$$\hat{S}_{\pm 1} e^{\mp\frac{\sigma}{S^2}\hat{S}_0} = e^{\frac{\sigma}{S^2}} e^{\mp\frac{\sigma}{S^2}\hat{S}_0} \hat{S}_{\pm 1}.$$

Here the magnitude of the ac field is supposed to be so large that the energy of a spin is either comparable to or higher than the thermal energy $kT$, i.e., $\xi \geq 1$, so that one is always faced with an intrinsically nonlinear problem which is solved as follows.

Now, the transformation of the evolution equation Eq. (4) for the reduced density matrix $\hat{\rho}$ into differential-recurrence equations for its *individual* matrix elements $\rho_{nm}$ may be accomplished because the diagonal entries $\rho_{mm}$ of the density matrix then decouple from the non-diagonal ones. Hence, only the former contribute to the longitudinal spin relaxation allowing a complete solution. Consequently, we have from Eq. (4) the following 3-term differential-recurrence equation for the *diagonal* entries $\rho_m = \rho_{mm}$, viz.

$$\tau_N \frac{d\rho_m(t)}{dt} = q_m^-(t)\rho_{m-1}(t) + q_m(t)\rho_m(t) + q_m^+(t)\rho_{m+1}(t), \qquad (5)$$

where $m = -S, -S+1, ..., S$, $\tau_N = (2D_\perp)^{-1}$ is the characteristic diffusion time and the time dependent coefficients $q_m(t)$ and $q_m^\pm(t)$ are

$$q_m(t) = -a_m^- e^{-(2m-1)\frac{\sigma}{2S^2} - \frac{\xi_0 + \xi\cos\omega t}{2S}} - a_m^+ e^{(2m+1)\frac{\sigma}{2S^2} + \frac{\xi_0 + \xi\cos\omega t}{2S}},$$

$$q_m^\pm(t) = a_m^\pm e^{\mp(2m\pm 1)\frac{\sigma}{2S^2} \mp \frac{\xi_0 + \xi\cos\omega t}{2S}},$$

$$a_m^\pm = \frac{(S \mp m)(S \pm m + 1)}{2}.$$

Since we are solely concerned with the ac response corresponding to the stationary state, which is independent of the initial conditions, in calculating the longitudinal component of the magnetization $\langle\hat{S}_Z\rangle(t)$ defined as

$$\langle\hat{S}_Z\rangle(t) = \sum_{m=-S}^{S} m\rho_m(t), \qquad (6)$$

we may seek the diagonal elements $\rho_m(t)$ as the Fourier series, viz.,

$$\rho_m(t) = \sum_{k=-\infty}^{\infty} \rho_m^k(\omega) e^{ik\omega t}. \qquad (7)$$

As is evident from Eqs. (6) and (7), $\langle\hat{S}_Z\rangle(t)$ is then rendered as a Fourier series, viz.,

$$\left\langle \hat{S}_Z \right\rangle(t) = \sum_{k=-\infty}^{\infty} S_Z^k(\omega)e^{ik\omega t}, \tag{8}$$

where the amplitudes $S_Z^k(\omega)$ are themselves given by the finite series

$$S_Z^k(\omega) = \sum_{m=-S}^{S} m\rho_m^k(\omega). \tag{9}$$

Next, the time dependent coefficients $q_m(t)$ and $q_m^\pm(t)$ in Eq. (5) can also be expanded into the Fourier series using the known Fourier-Bessel expansion [31]

$$e^{\pm\frac{\xi}{2S}\cos\omega t} = \sum_{k=-\infty}^{\infty} I_k\left(\pm\frac{\xi}{2S}\right)e^{ik\omega t}, \tag{10}$$

where $I_k(z)$ are the modified Bessel functions of the first kind [31]. Thus by direct substitution of Eq. (7) and the Fourier series for $q_m(t)$ and $q_m^\pm(t)$ into Eq. (5), we can derive a recurrence relation in $(k,m)$ between the Fourier coefficients $\rho_m^k(\omega)$, viz.,

$$ik\omega\tau_N \rho_m^k(\omega) = \sum_{k'=-\infty}^{\infty}\left\{ a_m^- e^{\frac{\sigma(2m-1)}{2S^2}+\frac{\xi_0}{2S}} I_{k-k'}\left(\frac{\xi}{2S}\right)\rho_{m-1}^{k'}(\omega) \right.$$
$$+ a_m^+ e^{-\frac{\sigma(2m+1)}{2S^2}-\frac{\xi_0}{2S}} I_{k-k'}\left(-\frac{\xi}{2S}\right)\rho_{m+1}^{k'}(\omega) \tag{11}$$
$$\left. -\left[ a_m^- e^{-\frac{\sigma(2m-1)}{2S^2}-\frac{\xi_0}{2S}} I_{k-k'}\left(-\frac{\xi}{2S}\right) + a_m^+ e^{\frac{\sigma(2m+1)}{2S^2}+\frac{\xi_0}{2S}} I_{k-k'}\left(\frac{\xi}{2S}\right) \right]\rho_m^{k'}(\omega) \right\}.$$

The recurrence relation Eq. (11) can be solved exactly for the Fourier amplitudes $\rho_m^k(\omega)$ via matrix continued fractions [25,32] (see Appendix B). Thus, having calculated $\rho_m^k(\omega)$, we have from Eq. (9) all the constituent frequency-dependent Fourier amplitudes $S_Z^k(\omega)$.

### III. LINEAR AND NONLINEAR DYNAMIC SUSCEPTIBILITIES

Initially, we treat the fundamental component of the magnetization $S_Z^1(\omega)$ describing the linear ac stationary response of a nanomagnet to a *vanishing ac driving field*, i.e., when the ac field parameter $\xi \to 0$. Here the normalized fundamental component $S_Z^1(\omega)/S_Z^1(0)$ yields the normalized *linear* dynamic susceptibility of the nanomagnet, viz.,

$$\frac{\chi(\omega)}{\chi} = \frac{S_Z^1(\omega)}{S_Z^1(0)}, \tag{12}$$

where $\chi$ is the static susceptibility defined as

$$\chi = \left\langle \hat{S}_Z^2 \right\rangle_0 - \left\langle \hat{S}_Z \right\rangle_0^2 = \sum_{m=-S}^{S} m^2 \rho_m^0 - \left( \sum_{m=-S}^{S} m\rho_m^0 \right)^2$$

with the matrix elements $\rho_m^0$ given by

$$\rho_m^0 = \frac{e^{\sigma m^2/S^2 + \xi_0 m/S}}{\sum_{m=-S}^{S} e^{\sigma m^2/S^2 + \xi_0 m/S}}, \tag{13}$$

and the angular brackets $\langle \ \rangle_0$ denote the equilibrium statistical averaging. The dynamic susceptibility $\chi(\omega)$ can equivalently be obtained via the Kubo relation [25,30]

$$\frac{\chi(\omega)}{\chi} = 1 - i\omega \tilde{C}(\omega), \tag{14}$$

where

$$\tilde{C}(\omega) = \int_0^\infty C(t) e^{-i\omega t} dt$$

is the one-sided Fourier transform of the normalized longitudinal equilibrium correlation function $C(t)$ given by

$$C(t) = \frac{1}{\beta \chi} \left\langle \int_0^\beta \left[ \hat{S}_Z(-i\lambda\hbar) - \langle \hat{S}_Z \rangle_0 \right] \left[ \hat{S}_Z(t) - \langle \hat{S}_Z \rangle_0 \right] d\lambda \right\rangle.$$

The correlation function $C(t)$ describes the longitudinal spin relaxation of a uniaxial nanomagnet after infinitesimally small changes in the magnitude of the dc field $\mathbf{H}_0$ [25,30]. In other words, $C(t)$ is a relaxation function describing the linear transient response after that a *small* probing field $\mathbf{H}$ having been applied in the distant past $t = -\infty$ parallel to the uniform dc field $\mathbf{H}_0$ is suddenly switched off at $t = 0$. In the low- ($\omega \to 0$) and high- ($\omega \to \infty$) frequency limits, we have from Eq. (14) [25]

$$\frac{\chi(\omega)}{\chi} = 1 - i\omega \tau_{cor} + ..., \quad \omega \to 0, \tag{15}$$

$$\frac{\chi(\omega)}{\chi} = \frac{1}{i\omega \tau_{ef}} + ..., \quad \omega \to \infty. \tag{16}$$

where $\tau_{cor}$ are $\tau_{ef}$ are, respectively, the integral and effective relaxation times given by

$$\tau_{cor} = \int_0^\infty C(t) dt \text{ and } \tau_{ef} = -\frac{C(0)}{\dot{C}(0)}.$$

We remark that the characteristic times $\tau_{cor}$ and $\tau_{ef}$ for a quantum uniaxial nanomagnet have been calculated by Garanin [13] and Garcia-Palacios and Zueco [16] thereby yielding analytic expressions for $\tau_{cor}$ and $\tau_{ef}$ even for more general models of spin-bath interactions than we have used here. Applied to the model to hand, their results become [19]

$$\tau_{cor} = \frac{2\tau_N}{\chi} \sum_{k=1-S}^{S} \frac{\left( \sum_{m=k}^{S} \left( m - \langle \hat{S}_Z \rangle_0 \right) \rho_m^0 \right)^2}{[S(S+1) - k(k-1)] \sqrt{\rho_k^0 \rho_{k-1}^0}} \tag{17}$$

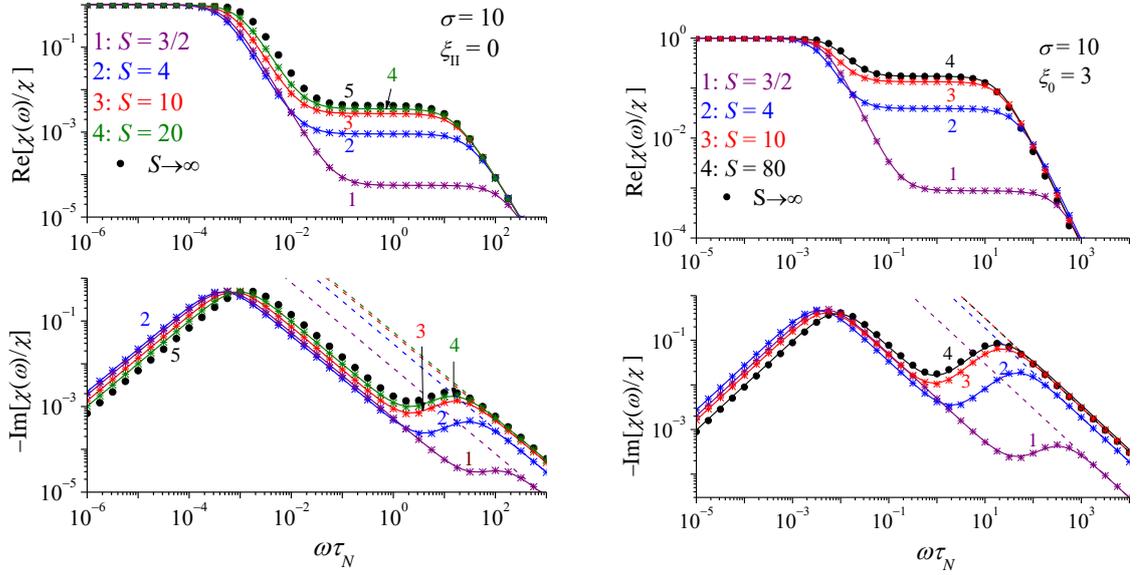

FIG. 1. (Color on line) Normalized linear susceptibility $\chi(\omega)/\chi$, Eq. (12), vs. $\omega\tau_N$ for the anisotropy parameter $\sigma=10$, the dc bias field parameter $\xi_0 = 0$ (left) and $\xi_0 = 3$ (right), and various spin numbers $S$. Asterisks: the two-mode approximation, Eq. (21). Dashed lines: the high-frequency asymptote, from Eqs. (16) and (18). Filled circles: the classical limit $S \to \infty$.

and

$$\tau_{\text{ef}} = \frac{2\chi\tau_N}{\sum_{k=1-S}^{S}[S(S+1)-k(k-1)]\sqrt{\rho_k^0 \rho_{k-1}^0}}. \qquad (18)$$

In Fig. 1, we plot the real and imaginary parts of the linear dynamic susceptibility $\chi(\omega)/\chi$ as calculated from the matrix continued fraction solution for zero dc field, $\xi_0 = 0$ (symmetrical wells) and for nonzero dc field, $\xi_0 = 3$ (asymmetrical wells). Two distinct bands appear in the magnetic loss spectrum $-\text{Im}[\chi(\omega)]$ and two corresponding dispersion regions occur in the spectrum of the real part of the susceptibility $\text{Re}[\chi(\omega)]$. The characteristic frequency and the half-width of this band are determined by the smallest *nonvanishing* eigenvalue $\lambda_1$ of the system matrix Eq. (C1) from Appendix C. The inverse of $\lambda_1$ determines the longest spin relaxation (or reversal) time $\tau = 1/\lambda_1$. Thus the reversal time $\tau$ can also be evaluated the frequency $\omega_{\text{max}}$ of the low frequency peak in the magnetic loss spectrum $-\text{Im}[\chi(\omega)]$, where it attains a maximum, and/or the half-width $\Delta\omega$ of the spectrum $\text{Re}[\chi(\omega)]$ via

$$\tau \approx \omega_{\text{max}}^{-1} \approx \Delta\omega^{-1}. \qquad (19)$$

Furthermore, the reversal time $\tau$ can be evaluated using Garanin's method [13] via the analytical equation [19]

$$\tau = \frac{2\tau_N}{\chi_\Delta} \sum_{k=-S}^{S-1} \frac{\left(\sum_{m=-S}^{k}\left(m-\langle\hat{S}_z\rangle_0\right)\rho_m^0\right)\left(\sum_{m=-S}^{k}[\mathrm{sgn}(m-m_b)-\Delta]\rho_m^0\right)}{[S(S+1)-k(k+1)]\sqrt{\rho_k^0 \rho_{k+1}^0}}, \qquad (20)$$

where $m_b$ is the quantum number corresponding to the top of the barrier with

$$\Delta = \sum_{m=-S}^{S} \mathrm{sgn}(m-m_b)\rho_m^0$$

and

$$\chi_\Delta = \sum_{m=-S}^{S} m\,\mathrm{sgn}(m-m_b)\rho_m^0 - \left(\sum_{m=-S}^{S} m\rho_m^0\right)\left(\sum_{m=-S}^{S} \mathrm{sgn}(m-m_b)\rho_m^0\right).$$

Comparison of $\tau$ as extracted from the spectra $\chi(\omega)$ via Eq. (19) with $\tau$ calculated independently via the smallest nonvanishing eigenvalue of the system matrix $\tau = \lambda_1^{-1}$ or via the analytic Eq. (20) shows that all methods yield the same results. Now, the second high-frequency band of $-\mathrm{Im}[\chi(\omega)]$ and the high-frequency dispersion region of $\mathrm{Re}[\chi(\omega)]$ are due to high-frequency "intrawell" modes corresponding to the near degenerate eigenvalues $\lambda_k$ ($k \geq 2$) of the system matrix Eq. (C1) from Appendix C. These individual "intrawell" modes are indistinguishable in the spectrum $-\mathrm{Im}[\chi(\omega)]$ appearing merely as a single high-frequency Lorentzian band. Thus, we may describe the behavior of $\chi(\omega)$ via a two-mode approximation, i.e., by supposing that $\chi(\omega)$ is given as a sum of two Lorentzians, viz., [16,22,25,33]

$$\frac{\chi(\omega)}{\chi} \approx \frac{1-\delta}{1+i\omega\tau} + \frac{\delta}{1+i\omega\tau_W}. \qquad (21)$$

Here $\tau_W$ is a characteristic relaxation time of the near degenerate high-frequency well modes and $\delta$ denotes a parameter characterizing their contribution to the susceptibility; they are defined as

$$\delta = \frac{\tau_{\mathrm{cor}}/\tau + (\tau-\tau_{\mathrm{cor}})/\tau_{\mathrm{ef}} - 1}{\tau_{\mathrm{cor}}/\tau + \tau/\tau_{\mathrm{ef}} - 2}, \qquad (22)$$

$$\tau_W = \frac{\tau_{\mathrm{cor}} - \tau}{1 - \tau/\tau_{\mathrm{ef}}}. \qquad (23)$$

The parameters $\delta$ and $\tau_W$ in Eqs. (21) and (22) have been determined by imposing the condition that the approximate two-mode Eq. (21) must obey the *exact* asymptotic Eqs. (15) and (16). In order to verify the accuracy of the two-mode approximation, we compare it in Fig. 1 with the matrix continued fraction solution. It is apparent from Fig. 1 that no *practical difference exists between the numerical solution and the two-mode approximation* (the maximum relative deviation between the corresponding curves does not exceed a few percent). In the classical limit, $S \to \infty$, the Hamiltonian Eq. (1) corresponds to a normalized free energy $V$ given by

$$\beta V(\vartheta) = -\sigma\cos^2\vartheta - \xi_0\cos\vartheta. \qquad (24)$$

Results for this classical limit are also shown in Fig. 1 for comparison. Our conclusions mirror those of Garcia-Palacios and Zueco [16] who have also shown that the two-mode approximation, which was originally developed for classical systems [33], accurately describes the linear response of quantum nanomagnets.

Turning our attention to the nonlinear response, we see that in strong ac fields, pronounced nonlinear effects occur as the amplitude of the ac field increases (see Figs. 2 and 3). The fundamental component $S_Z^1(\omega)/S_Z^1(0)$ is shown in Fig. 2 for various ac and dc field parameters. As in the linear response, two distinct absorption bands again appear in the spectrum of $-\text{Im}[S_Z^1(\omega)/S_Z^1(0)]$ and two corresponding dispersion regions occur in the spectrum of $\text{Re}[S_Z^1(\omega)/S_Z^1(0)]$, see Fig. 2. However, due to the pronounced nonlinear effects, the low-frequency parts of $-\text{Im}[S_Z^1(\omega)/S_Z^1(0)]$ and $\text{Re}[S_Z^1(\omega)/S_Z^1(0)]$ may no longer be approximated by a *single* Lorentzian. Nevertheless, the maximum loss frequency $\omega_{\max}$ and/or the half-width $\Delta\omega$ of the spectrum $\text{Re}[S_Z^1(\omega)/S_Z^1(0)]$ may still be used to estimate an effective reversal time $\tau$ as defined by Eq. (19). The behavior of the low-frequency peak of $-\text{Im}[S_Z^1(\omega)/S_Z^1(0)]$ as *a function of the ac field amplitude* crucially depends on whether or not a dc field is applied. For strong dc bias, $\xi_0 > 1$ (see Fig. 2), the low-frequency peak shifts to lower frequencies reaching a maximum at $\xi \sim \xi_0$ thereafter shifting to higher frequencies with increasing $\xi_0$. In other words, as the dc field increases, the reversal time of the spin *initially increases* and having attained its maximum at some critical value $\xi \sim \xi_0$ thereafter decreases. This behavior agrees with that observed in the classical case [16,33]. For zero dc bias, $\xi_0 = 0$, the low-frequency peak shifts to higher frequencies with increasing $\xi$. Now, a striking feature of the nonlinear response is that the effective reversal time $\tau$ may also be evaluated from either the spectrum of the (now) frequency dependent dc component $S_Z^0(\omega)$ (only for nonzero dc bias, $\xi_0 \neq 0$) or those of the higher order harmonics $S_Z^k(\omega)$ with $k > 1$ because the low-frequency parts of these spectra are, like the spectra of the fundamental, themselves dominated by overbarrier relaxation processes. For illustration, the real and imaginary parts of the second and third harmonic components $S_Z^2(\omega)/S_Z^2(0)$ and $S_Z^3(\omega)/S_Z^3(0)$ are shown in Fig. 3. Like the fundamental, the behavior of both $-\text{Im}\left[S_Z^2(\omega)/S_Z^2(0)\right]$ and $-\text{Im}\left[S_Z^3(\omega)/S_Z^3(0)\right]$ depends on whether or not a dc field is applied. For zero and weak dc bias field $\xi_0 < 0.5$, the low-frequency peak shifts monotonically to higher frequencies. For strong dc bias field, $\xi_0 > 1$, on the other hand the low-frequency peak shifts to lower frequencies reaching a maximum at $\xi \sim \xi_0$ thereafter decreasing with increasing $\xi$.

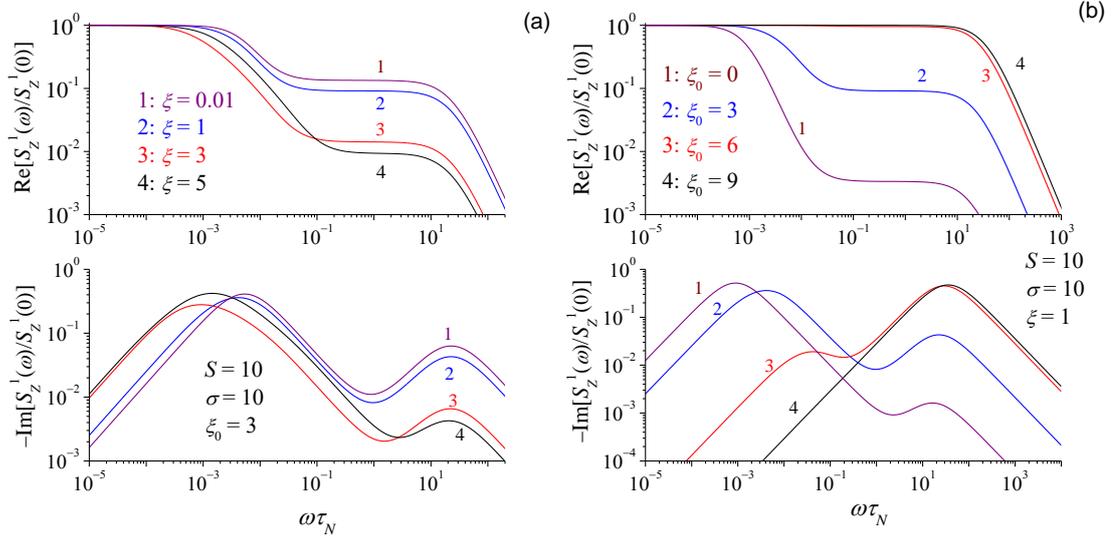

FIG. 2. (Color on line) The real and imaginary parts of the normalized fundamental component $S_Z^1(\omega)/S_Z^1(0)$ vs. $\omega\tau_N$ (a) for various values of the applied ac stimulus $\xi = 0.01$ (linear response), 1, 3, 5 and the dc field parameter $\xi_0 = 3$ and (b) for various dc field parameters $\xi_0$ and $\xi = 1$; the spin number $S = 10$ and anisotropy parameter $\sigma = 10$.

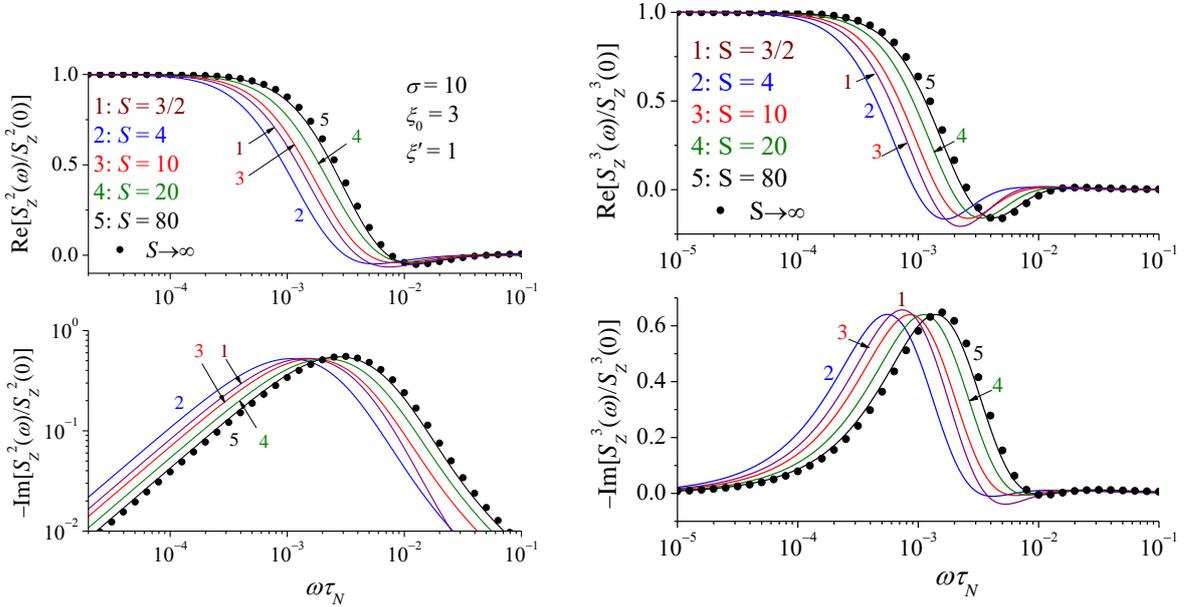

FIG. 3. (Color on line) The real and imaginary parts of the normalized second and third harmonic components $S_Z^2(\omega)/S_Z^2(0)$ and $S_Z^3(\omega)/S_Z^3(0)$ vs. $\omega\tau_N$ for anisotropy parameter $\sigma = 10$, the dc field parameter $\xi_0 = 3$, the ac field parameter $\xi = 1$, and various spin numbers $S$. Filled circles: the classical limit.

## IV. DYNAMIC MAGNETIC HYSTERESIS

Studies of DMH in magnetic nanoparticles having been initiated by Ignatchenko and Gekht [9] were later extended in many other investigations (see, e.g., Refs. [10-12]). Like the classical case, DMH loops for quantum nanomagnets represent a parametric plot of the normalized magnetization as a function of the normalized ac field, i.e.,

$$m(t) = \langle \hat{S}_Z \rangle(t)/S \text{ vs. } h(t) = H(t)/H = \cos \omega t. \tag{25}$$

Just as the classical case [11,12,25], the normalized area of the DMH loop $A_n$, which is the energy loss per particle over one cycle of the ac field, is given by

$$A_n = \frac{1}{4} \oint m(t) dh(t) = -\frac{\pi}{2S} \text{Im}\left[ S_Z^1(\omega) \right]. \tag{26}$$

In Figs. 4-7, we show the effects of ac and dc bias magnetic fields on the DMH loops in a uniaxial nanomagnet for various $S$. For a weak ac field, $\xi \to 0$, and low frequencies, $\omega \tau \leq 1$, the DMH loops are ellipses with normalized area $A_n$ given by Eq. (26); the behavior of $A_n \sim -\text{Im}\left[ S_Z^1(\omega) \right] \sim \chi''(\omega)$ being similar [cf. Eq. (26)] to that of the magnetic loss $\chi''(\omega)$ (see Fig. 1). Indeed, the two-mode approximation Eq. (21) for the susceptibility implies that the overall relaxation process consists of two distinct entities, namely, the slow thermally activated overbarrier (interwell) process and the fast (intrawell) relaxation in the wells. Now, at low frequencies and for large barriers between the wells, only the first term on the right side in Eq. (21) need be considered. Furthermore, for weak dc bias fields, $\xi_0/(2\sigma) \ll 1$, the approximation $\delta \approx 1$ may always be used in Eq. (21) so that $m(t) = \langle \hat{S}_Z \rangle(t)/S$ can be given by the simple linear response formula [12,25]

$$m(t) \approx \frac{1}{S}\langle \hat{S}_Z \rangle_0 + \frac{\chi \xi}{S} \frac{\cos \omega t + \omega \tau \sin \omega t}{1 + \omega^2 \tau^2}. \tag{27}$$

If we introduce the variables $x$ and $y$ defined as

$$x = \cos \omega t \text{ and } y = \frac{1}{\chi \xi}\left[ Sm(t) - \langle \hat{S}_Z \rangle_0 \right],$$

we then can conclude from Eqs. (25) and (27) that in the linear response approximation, a low frequency DMH loop in the $(x, y)$ plane is an ellipse, namely, [12]

$$x^2 + \frac{1}{\omega^2 \tau^2}\left[ (1 + \omega^2 \tau^2) y - x \right]^2 = 1. \tag{28}$$

This ellipse is centered at $(0,0)$ and its major axis is inclined to the $x$-axis at an angle $\frac{1}{2}\arctan\left[ 2/(\omega \tau)^2 \right]$ [12].

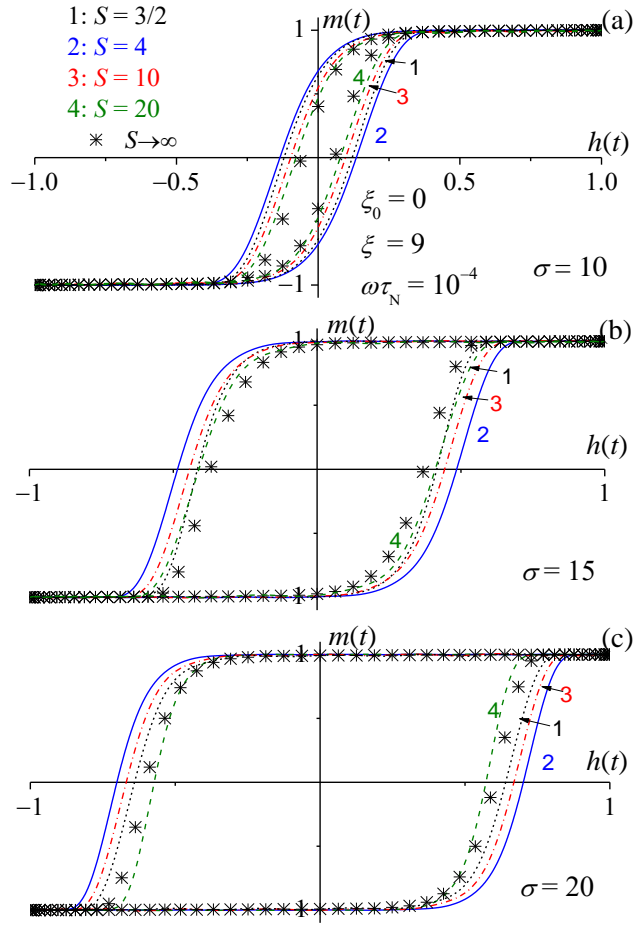

FIG. 4. (Color on line) DMH loops [$m(t) = \langle \hat{S}_Z \rangle(t)/S$ vs. $h(t) = \cos \omega t$] for various anisotropy parameters $\sigma = 10$ (a), 15 (b), 20 (c) and various spin numbers $S = 3/2$ (1: short-dashed lines), 4 (2: solid lines), 10 (3: dashed-dotted lines), 20 (4: dashed lines), and $\infty$ (asterisks) at $\omega \tau_N = 10^{-4}$, $\xi_0 = 0$, and $\xi = 9$.

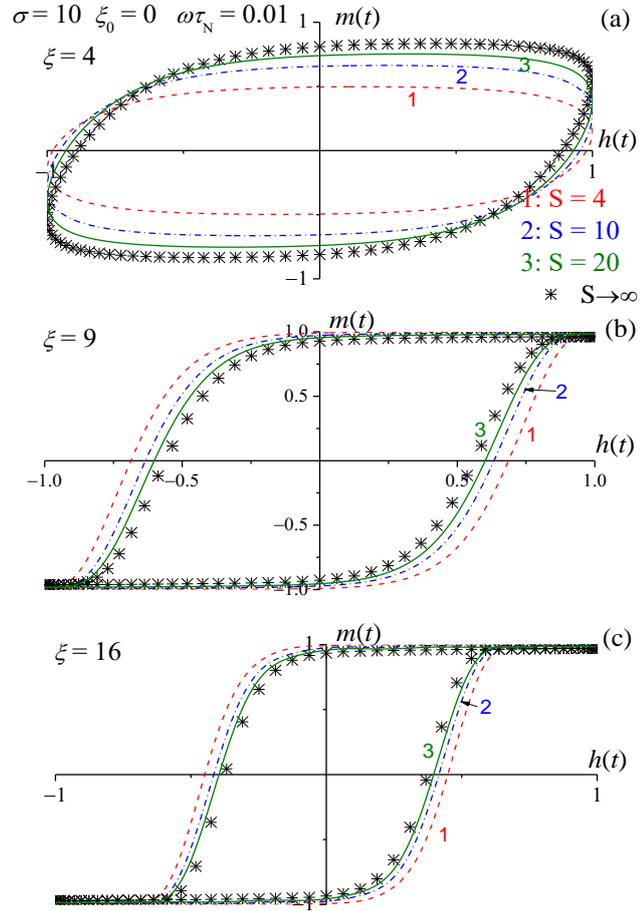

FIG. 5. (Color on line) DMH loops for various ac field parameters $\xi = 4$ (a), 9 (b), 16 (c) and various spin numbers $S = 4$ (1: dashed lines), 10 (2: dashed-dotted lines), 20 (3: solid lines), and $\infty$ (asterisks) at $\sigma = 10$, $\xi_0 = 0$, and $\omega\tau_N = 10^{-2}$.

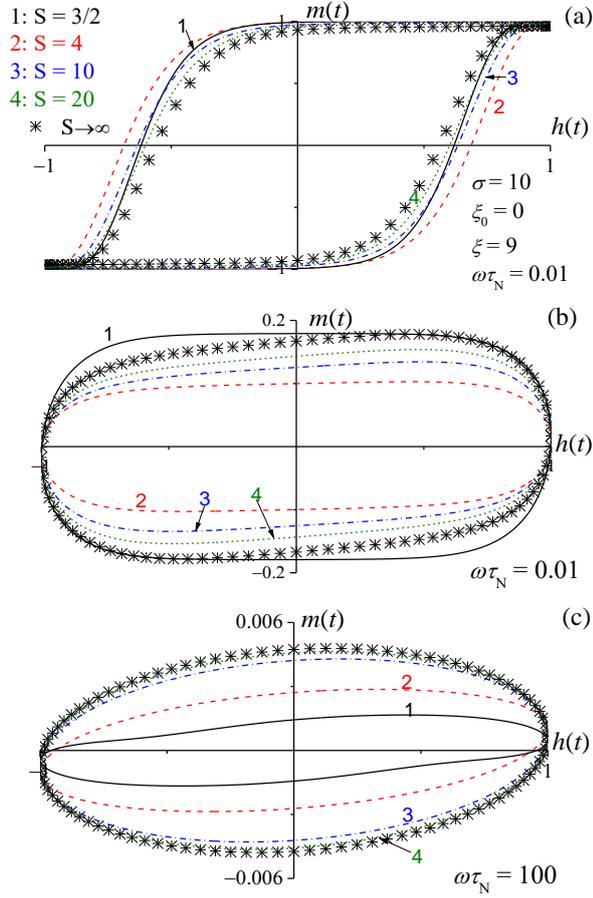

FIG. 6. (Color on line) DMH loops for various frequencies $\omega\tau_N = 10^{-2}$ (a), 1 (b), $10^2$ (c) and various spin numbers $S = 3/2$ (1: solid lines), 4 (2: dashed lines), 10 (3: dashed-dotted lines), 20 (4: short-dashed lines), and $\infty$ (asterisks) at $\sigma = 10$, $\xi_0 = 0$, and $\xi = 9$.

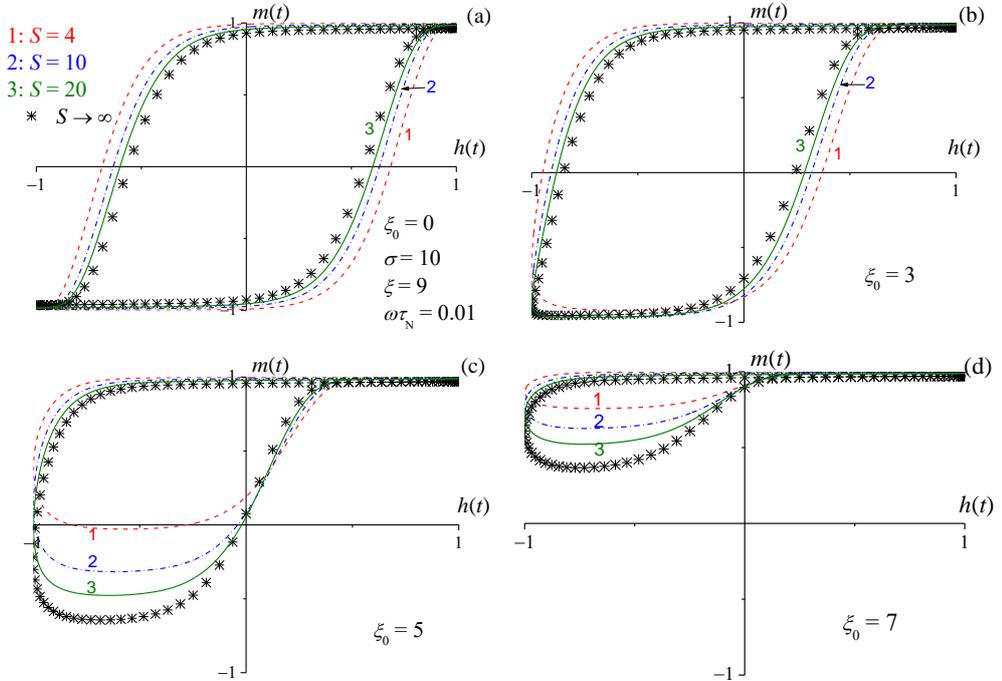

FIG. 7. (Color on line) DMH loops for various dc field parameters $\xi_0 = 0$ (a), 3 (b), 5 (c), 7 (d) and various spin numbers $S = 4$ (1: dashed lines), 10 (2: dashed-dotted lines), 20 (3: solid lines), and $\infty$ (asterisks) at $\xi = 9$, $\sigma = 10$, and $\omega\tau_N = 10^{-2}$.

For *moderate* ac fields, $0.5 < \xi \leq 1$, the DMH loops still have approximately an ellipsoidal shape implying that only a few harmonics actually contribute to the weakly nonlinear response. In contrast, in *strong* ac fields, $\xi > 1$, the shape of DMH loops alters substantially and so the normalized area $A_n$ now exhibiting a pronounced dependence on the frequency $\omega$, the ac and dc bias field amplitudes $\xi$ and $\xi_0$ as well as on the anisotropy parameter $\sigma$ and the spin number $S$ (see Figs. 4-7). In this regime, the external ac field is able to saturate the magnetization as well as being to induce its inversion (i.e., switching between the directions of the easy axis). In Figs. 4 and 5, we plot the DMH loops for various spin numbers $S$ and anisotropy ($\sigma$) and ac field ($\xi$) parameters exemplifying how DMH loops alter as these parameters vary. Clearly, the re-magnetization time is highly sensitive to variations of these parameters. For example, with a strong ac driving field, the Arrhenius dependence of the reversal time on temperature $\log \tau \propto 1/T$, which accurately accounts for the linear response regime, is modified because the strong ac field intervenes so drastically reducing the effective response time of the nanomagnet. Thus, the nonlinear behavior facilitates re-magnetization regimes, which are never attainable with weak ac fields. The reason being that the dc bias field under the appropriate conditions *efficiently tunes* this effect by either *enhancing* or *blocking* the action of the strong ac field. The *pronounced* frequency dependence of the loops is highlighted in Fig. 6 for various $S$. At low frequencies, the field changes are *quasi-adiabatic*, so that the magnetization reverses due to the *cooperative* shuttling action of thermal agitation combined with the ac field. The dc bias field effects on the DMH are illustrated in Fig. 7 showing the changes in the DMH caused by varying $\xi_0$ for various spin numbers $S$. In order to understand the effect of the dc bias field on the loop area, one must first recall that the magnetic reversal time depends on the actual value of the applied field. Under the conditions of Fig. 7, the *positive limiting* (saturation) value of $m(t) \to 1$ corresponds to a total field $H_0 + H$ thus *favoring* the magnetization relaxation to the positive saturation value $m(t) \to 1$. However, for *negative* $h(t)$, the total field $H_0 - H$ is much weaker and so cannot induce relaxation to the negative saturation value $m(t) \to -1$. Therefore, the "center of area" of the loop moves upwards. In the classical limit, $S \to \infty$, our results concur with those for classical uniaxial nanomagnets [12].

The temperature dependence of the DMH is governed by the anisotropy (inverse temperature) parameter $\sigma \propto 1/T$. The normalized DMH area $A_n$ as a function of $\sigma^{-1}$ is shown in Fig. 8 for various $S$ showing that the tuning action of the dc bias field described above is effective over a certain temperature interval. This conclusion once again indicates that the relaxation of the magnetization is mostly caused by thermal fluctuations, implying that the magnetic response time retains a strong temperature dependence. The normalized area as a function of the frequency $\omega$ and ac field parameter $\xi/(2\sigma)$ is shown in Figs. 9 and 10, respectively. Clearly $A_n$ can invariably be represented as a nonmonotonic curve with a maximum the position of which is determined by $S$ as well as by the other model parameters. The peak in $A_n$ (Fig. 9) is caused by the field-induced modifications of the reversal time as strongly tuned by the dc bias field. As in Fig. 9, variation of the dc field strength shifts the frequency, where the maximum is attained, by several orders of magnitude. The normalized loop area presented in Fig. 10 illustrates the dependence of $A_n$ on the ac field amplitude, which is similar to that of classical nanomagnets [12].

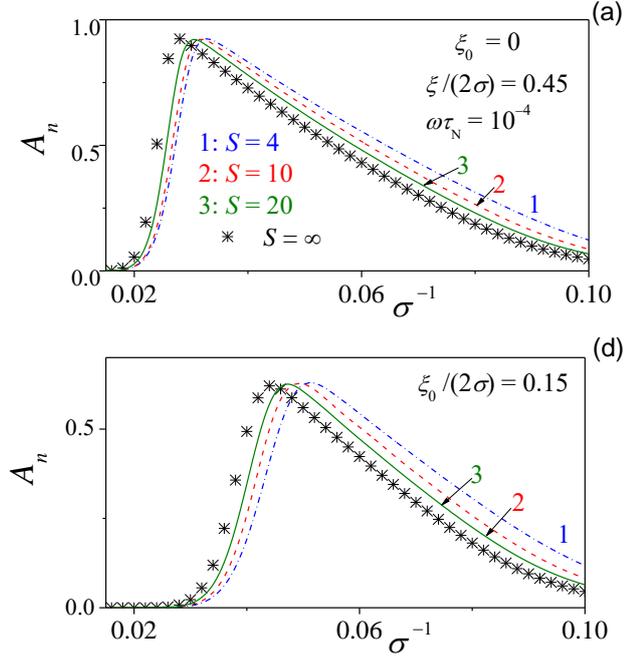

FIG. 8. (Color on line) Normalized area of the DMH loop $A_n$ vs. the dimensionless temperature $\sigma^{-1}$ under variation of the dc bias field parameter $h_0 = \xi_0/(2\sigma) = 0$ (a) and 0.15 (b) for various spin numbers $S = 4$ (dashed-dotted lines), 10 (dashed lines), 20 (solid lines), and $\infty$ (asterisks) at the frequency $\omega\tau_N = 10^{-4}$ and the ac field amplitude $\xi/(2\sigma) = 0.45$.

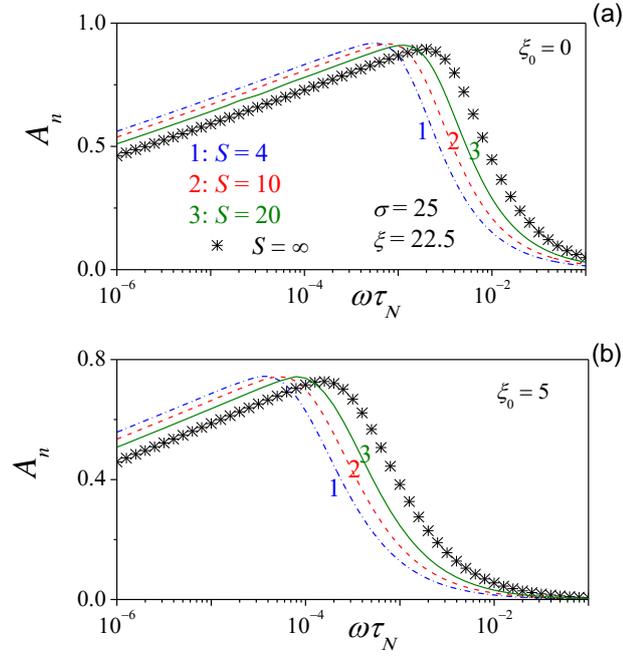

FIG. 9. (Color on line) Normalized area of the DMH loop $A_n$ vs. $\omega\tau_N$ under variation of the dc bias field $\xi_0 = 0$ (a) and 5 (b) for various spin numbers $S = 4$ (dashed-dotted lines), 10 (dashed lines), 20 (solid lines), and $\infty$ (asterisks). The anisotropy parameter $\sigma = 25$ and the ac field parameter $\xi/(2\sigma) = 0.45$.

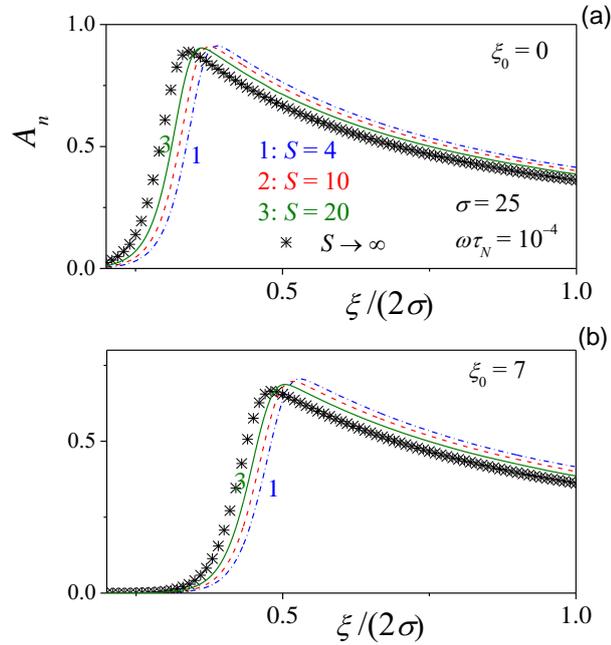

FIG. 10. (Color on line) Normalized area of the DMH loop $A_n$ vs. $\xi/(2\sigma)$ under variation of the bias field parameter $\xi_0 = 0$ (a), 2.5 (b), 5 (c), and 7 (d) for various spin numbers $S = 4$ (dashed-dotted lines), 10 (dashed lines), 20 (solid lines), and $\infty$ (asterisks). The anisotropy parameter $\sigma = 25$ and frequency $\omega\tau_N = 10^{-4}$.

## V. CONCLUSIONS

We have studied the nonlinear ac stationary response of uniaxial nanomagnets with arbitrary spin number *S* subjected to superimposed ac and dc magnetic fields in the high temperature and weak spin-bath coupling limit. The nonlinear dynamic susceptibility and DMH in such nanomagnets has been treated without any *a priori* assumptions regarding the magnetizing field strength and the spin number *S*. In general, it appears that a small (in comparison with the internal anisotropy field) bias dc field can profoundly affect the nonlinear dynamic susceptibility and shape of the DMH loops in nanomagnets accompanied by a strong dependence on *S*. The overall conclusion is that just as in linear response [16,19], one may determine the transition from quantum elementary spin relaxation to that pertaining to a giant spin as a function of the spin number *S* yielding explicitly the evolution of the nonlinear ac stationary response and DMH from that of molecular magnets (*S* ~ 10) to nanoclusters (*S* ~ 100), and to classical superparamagnets (*S* > 1000). For linear response, the results obtained entirely agree with those given in Ref. [19] while in the classical limit ($S \to \infty$), the solutions obtained via the evolution equation for the density matrix reduce to those yielded by the Fokker-Planck equation for the orientation distribution function of classical (macro)spins [25,26]. Hence, the results indicate that quantum effects in the nonlinear spin relaxation can be treated in a manner linking directly to the magnetization relaxation in nanomagnets with classical superparamagnetic behavior. Here we have only considered the nonlinear dynamic susceptibility and DMH of uniaxial nanomagnets in the simplest configuration, i.e., where the ac and dc magnetic fields are applied along the easy axis of the magnetization. However, the calculation may, in principle, be generalized to other interesting cases such as arbitrary directions of applied fields and nonaxially symmetric anisotropies (cubic, biaxial, etc.).

## ACKNOWLEDGMENTS

This publication has emanated from research conducted with the financial support of FP7-PEOPLE-Marie Curie Actions - International Research Staff Exchange Scheme (Project No. 295196 DMH) for financial support.

## Appendix A: Collision kernel in the high temperature limit

To derive Eq. (3), we follow Hubbard [28] who considered the general case of the time-dependent Hamiltonian $\hat{H}_S = \hat{H}_S(t)$. The collision kernel used by Hubbard is (in our notation)

$$\text{St}(\hat{\rho}) = \sum_{\mu=-1}^{1} \sum_{r} (-1)^\mu e^{i\omega_r^{-\mu} t} D_\mu(\omega_r^{-\mu}) \left\{ e^{\beta\hbar\omega_r^{-\mu}/2} \left[ \hat{S}_\mu, \hat{\rho} \hat{U}^{-1}(t) \hat{S}_{-\mu}^r \hat{U}(t) \right] \right.$$
$$\left. + e^{-\beta\hbar\omega_r^{-\mu}/2} \left[ \hat{U}^{-1}(t) \hat{S}_{-\mu}^r \hat{U}(t) \hat{\rho}, \hat{S}_\mu \right] \right\}, \quad \text{(A1)}$$

where $\hat{S}_{\mu'}^r$ are the coefficients in the series expansion of the time-dependent spin operators $\hat{S}_{\mu'}(t) = \hat{U}(t)\hat{S}_{\mu'}\hat{U}^{-1}(t)$, namely,

$$\hat{S}_{\mu'}(t) = \sum_r \hat{S}_{\mu'}^r e^{i\omega_r^{\mu'} t}, \tag{A2}$$

where $\omega_r^{\mu'}$ represents a parameter, while the operator $\hat{U}(t)$ is defined as

$$\hat{U}(t) = e^{\frac{i}{\hbar}\int_0^t \hat{H}_S(t')dt'}, \tag{A3}$$

and $D_\mu(\omega)$ is the correlation function of the bath written in the frequency domain as

$$D_\mu = \tilde{C}_\mu^{sym}(\omega)\operatorname{sech}(\beta\hbar\omega/2), \tag{A4}$$

with the symmetrized spectral density $\tilde{C}_\mu^{sym}(\omega) = \left[\tilde{C}_{\mu,-\mu}(-\omega) + \tilde{C}_{\mu,-\mu}^*(\omega)\right]/2$ which determines the spectrum of the *symmetrized* bath correlation functions. Then by reconverting the result to operator form [see Eq. (A2)], we have for the collision kernel

$$\begin{aligned}\operatorname{St}(\hat{\rho}) &= \sum_{\mu=-1}^{1}\sum_r (-1)^\mu D_\mu e^{i\omega_r^{-\mu}t}\\ &\times\left(e^{\beta\hbar\omega_r^{-\mu}/2}\left[\hat{S}_\mu, \hat{\rho}\hat{U}^{-1}(t)\hat{S}_{-\mu}^r\hat{U}(t)\right] + e^{-\beta\hbar\omega_r^{-\mu}/2}\left[\hat{U}^{-1}(t)\hat{S}_{-\mu}^r\hat{U}(t)\hat{\rho}, \hat{S}_\mu\right]\right)\\ &= \sum_{\mu=-1}^{1}(-1)^\mu D_\mu\Big(\left[\hat{S}_\mu, \hat{\rho}\hat{U}^{-1}(t)\hat{U}(t-i\beta\hbar/2)\hat{S}_{-\mu}\hat{U}^{-1}(t-i\beta\hbar/2)U(t)\right]\\ &\quad + \left[\hat{U}^{-1}(t)\hat{U}(t+i\beta\hbar/2)\hat{S}_{-\mu}\hat{U}^{-1}(t+i\beta\hbar/2)\hat{U}(t)\hat{\rho}, \hat{S}_\mu\right]\Big).\end{aligned} \tag{A5}$$

Next, we consider typical products like $\hat{U}^{-1}(t)\hat{U}(t\pm i\beta\hbar/2)$ e.g. for the Hamiltonian $\hat{H}_S(t)$ given by

$$\hat{U}^{-1}(t)\hat{U}(t\pm i\beta\hbar/2) = e^{\frac{i}{\hbar}\int_t^{t\pm i\beta\hbar/2}\hat{H}_S(t')dt'}. \tag{A6}$$

We have for the integral

$$\frac{i}{\hbar}\int_t^{t\pm i\beta\hbar/2}\hat{H}_S(t')dt' = \mp\frac{\beta}{2}\hat{H}_S(t). \tag{A7}$$

Here we have made an approximation valid in the high temperature limit $\beta\hbar\omega \ll 1$ that is to say for small increments $\Delta t \sim \beta\hbar/2 \ll 1$ in Eq. (A7), we suppose that the operator $\hat{H}_S(t)$ does not alter significantly during the small time $\Delta t$. Thus, we can simply take the value of that operator value at time $t$ and consequently may place it outside the integral. By treating in like manner all other such time-dependent functions in Eq. (A5), we have the Hubbard form of the collision kernel Eq. (A5) with time dependent Hamiltonian $\hat{H}_S(t)$ which in the high temperature limit simplifies to Eq. (3). The form of the collision kernel given by Eq. (3) corresponds to the high temperature limit and short correlation time of the Markovian approximation. Use of the *symmetrized* collision kernel Eq. (3) is essential, because only this form ensures the absence of the even harmonics in the magnetization nonlinear response for symmetric uniaxial Hamiltonians like $\beta \hat{H}_S(t) = -\sigma \hat{S}_Z^2 / S^2$.

## Appendix B: Matrix continued fraction solution of Eq. (11)

On introducing the frequency-dependent column vector

$$\boldsymbol{\rho}_n = \begin{pmatrix} \vdots \\ \rho_n^{-1}(\omega) \\ \rho_n^{0}(\omega) \\ \rho_n^{1}(\omega) \\ \vdots \end{pmatrix}, \quad (B1)$$

$(n = m + S)$, we then have a homogeneous matrix three-term recurrence equation between column vectors $\boldsymbol{\rho}_n$, namely,

$$\mathbf{Q}_n^- \boldsymbol{\rho}_{n-1} + \mathbf{Q}_n \boldsymbol{\rho}_n + \mathbf{Q}_n^+ \boldsymbol{\rho}_{n+1} = \mathbf{0}, \quad (B2)$$

where the matrix elements of the infinite matrices $\mathbf{Q}_n$ and $\mathbf{Q}_n^\pm$ are given by

$$\left[\mathbf{Q}_n\right]_{kk'} = -i\omega \tau_N k \delta_{kk'} - a_n^+ e^{(2n-2S+1)\frac{\sigma}{S^2} + \frac{\xi_0}{S}} I_{k-k'}\left(\frac{\xi}{2S}\right) - a_n^- e^{-(2n-2S-1)\frac{\sigma}{2S^2} - \frac{\xi_0}{2S}} I_{k-k'}\left(-\frac{\xi}{2S}\right),$$

$$\left[\mathbf{Q}_n^\pm\right]_{kk'} = a_n^\pm e^{\mp(2n-2S\pm1)\frac{\sigma}{2S^2} \mp \frac{\xi_0}{2S}} I_{k-k'}\left(\mp \frac{\xi}{2S}\right).$$

However a nontrivial solution of the homogeneous Eq. (B2) exists because according to the general method of solution of three-term recurrence relations [25,32], all *higher order* column vectors $\boldsymbol{\rho}_n$ defined by Eq. (B1) can always be expressed in terms of the *lowest order* vector column $\boldsymbol{\rho}_0$ via the products

$$\boldsymbol{\rho}_n = \mathbf{S}_n \mathbf{S}_{n-1} \ldots \mathbf{S}_1 \boldsymbol{\rho}_0, \quad (B3)$$

where the $\mathbf{S}_m$ are finite matrix continued fractions defined by the matrix recurrence relation

$$\mathbf{S}_m = \left[-\mathbf{Q}_m - \mathbf{Q}_m^+ \mathbf{S}_{m+1}\right]^{-1} \mathbf{Q}_m^-.$$

Now the zero-order column vector $\boldsymbol{\rho}_0$ itself can be found from the normalization condition for the density matrix elements, viz.,

$$\sum_{n=0}^{2S} \rho_n(t) = \sum_{k=-\infty}^{\infty} \left(\sum_{n=0}^{2S} \rho_n^k(\omega)\right) e^{i\omega k t} = 1 \quad (B4)$$

thereby immediately yielding an *inhomogeneous* equation for $\boldsymbol{\rho}_0$, viz.,

$$\sum_{n=0}^{2S} \boldsymbol{\rho}_n = \mathbf{C}\boldsymbol{\rho}_0 = \mathbf{v}, \quad (B5)$$

where the matrix $\mathbf{C}$ is given by

$$\mathbf{C} = \mathbf{I} + \mathbf{S}_1 + \mathbf{S}_2\mathbf{S}_1 + ... + \mathbf{S}_{2S}...\mathbf{S}_2\mathbf{S}_1, \tag{B6}$$

$\mathbf{I}$ is the unit matrix, and the infinite column vector $\mathbf{v}$ has only one nonvanishing elements $v_k = \delta_{k0}$, $-\infty < k < \infty$. Consequently, we have for the zero-order column vector $\boldsymbol{\rho}_0$:

$$\boldsymbol{\rho}_0 = \mathbf{C}^{-1}\mathbf{v}. \tag{B7}$$

Having calculated all the $\boldsymbol{\rho}_0$, we can determine via Eq. (B3) the other column vectors $\boldsymbol{\rho}_n$ as

$$\boldsymbol{\rho}_n = \mathbf{S}_n\mathbf{S}_{n-1}...\mathbf{S}_1\mathbf{C}^{-1}\mathbf{v} \tag{B8}$$

and thus we can evaluate all the $S_Z^k(\omega)$ from Eq. (9) yielding the nonlinear stationary ac response of a uniaxial nanomagnet.

## Appendix C: Evaluation of the longest relaxation time $\tau$

In the absence of the ac driving field, i.e., $\xi = 0$, the recurrence relation Eq. (5) can be written in the *homogeneous* matrix form

$$\dot{\mathbf{F}}(t) = \mathbf{\Pi} \cdot \mathbf{F}(t),$$

where the column vector $\mathbf{F}(t)$ and the tridiagonal system matrix $\mathbf{\Pi}$ are

$$\mathbf{F}(t) = \begin{pmatrix} \rho_0(t) \\ \rho_1(t) \\ \vdots \\ \rho_{2S}(t) \end{pmatrix},$$

$$\mathbf{\Pi} = \frac{1}{\tau_N} \begin{pmatrix} p_0 & p_0^+ & 0 & \cdots & 0 \\ p_1^- & p_1 & p_1^+ & \cdots & \vdots \\ \vdots & \vdots & \vdots & \ddots & p_{2S-1}^+ \\ 0 & \cdots & 0 & p_{2S}^- & p_{2S} \end{pmatrix} \tag{C1}$$

with matrix elements

$$p_n = -\frac{n(2S-n+1)}{2}e^{-(2n-2S+1)\frac{\sigma}{2S^2}-\frac{\xi_0}{2S}} - \frac{(n+1)(2S-n)}{2}e^{(2n-2S+1)\frac{\sigma}{2S^2}+\frac{\xi_{II}}{2S}},$$

$$p_n^+ = \frac{1}{2}(2S-n)(n+1)e^{-(2n-2S-1)\frac{\sigma}{2S^2}-\frac{\xi_0}{2S}},$$

$$p_n^- = \frac{n}{2}(2S-n+1)e^{(2n-2S-1)\frac{\sigma}{2S^2}+\frac{\xi_0}{2S}}.$$

(these matrix elements are obtained from coefficients $q_m(t)$ and $q_m^\pm(t)$ in Eq. (5) by introducing a new index $n$ defined as $n = m + S$). The secular equation, which determines all the eigenvalues, is as usual

$$\det(\mathbf{\Pi} - \lambda\mathbf{I}) = 0. \tag{C2}$$

Now the left-hand side of Eq. (C2) represents a polynomial of the order $2S+1$, viz.,

$$\left(k_{2S+1}\lambda^{2S} + k_{2S}\lambda^{2S-1} + \ldots + k_2\lambda + k_1\right)\lambda = 0, \tag{C3}$$

where

$$k_1 = -\sum_{i=0}^{2S} M_i^i, \quad k_2 = \sum_{i=0}^{2S-1}\sum_{j=i+1}^{2S} M_{ij}^{ij},$$

and so on and we have used the fact that $\det(\mathbf{\Pi}) = 0$. Here the $M_i^{i'}$ are the first minors of the matrix $\mathbf{\Pi}$, which are the determinants of the square matrices as reduced from $\mathbf{\Pi}$ by removing the $i$th row and the $i'$th column of $\mathbf{\Pi}$ while the $M_{ij}^{i'j'}$ are the minors of the matrix $\mathbf{\Pi}$, which are in turn the determinants of the square matrix as reduced from $\mathbf{\Pi}$ by removing *two* (the $i$th and the $j$th) of its rows and *two* (the $i'$th and the $j'$th) columns. Now in the high barrier approximation when $\lambda_1 \ll 1$, that quantity can be evaluated analytically by neglecting all higher powers $\lambda^n$ with $n > 2$ in the secular Eq. (C3). Thus, we have from that equation

$$\lambda_1 \approx -\frac{k_1}{k_2}. \tag{C4}$$

However, Eq. (C4) can be equivalently written in matrix form as

$$\lambda_1 \approx \frac{\text{Tr}(\mathbf{M}^{(1)})}{\text{Tr}(\mathbf{M}^{(2)})}, \tag{C5}$$

where $\mathbf{M}^{(1)}$ is the matrix formed from all the first minors

$$\mathbf{M}^{(1)} = \begin{pmatrix} M_{2S}^{2S} & M_{2S}^{2S-1} & \cdots & M_{2S}^{0} \\ M_{2S-1}^{2S} & M_{2S-1}^{2S-1} & \cdots & M_{2S-1}^{0} \\ \vdots & \vdots & \ddots & \vdots \\ M_{0}^{2S} & M_{0}^{2S-1} & \cdots & M_{0}^{0} \end{pmatrix},$$

and the matrix $\mathbf{M}^{(2)}$ contains all the other $M_{ij}^{i'j'}$ minors

$$\mathbf{M}^{(2)} = \begin{pmatrix} M_{2S,2S-1}^{2S,2S-1} & M_{2S,2S-1}^{2S,2S-2} & \cdots & M_{2S,2S-1}^{0,0} \\ M_{2S,2S-2}^{2S,2S-1} & M_{2S,2S-2}^{2S,2S-2} & \cdots & M_{2S,2S-2}^{0,0} \\ \vdots & \vdots & \ddots & \vdots \\ M_{0,0}^{2S,2S-1} & M_{0,0}^{2S,2S-2} & \cdots & M_{0,0}^{0,0} \end{pmatrix}.$$

The matrices $\mathbf{M}^{(1)}$ and $\mathbf{M}^{(2)}$ have, respectively, dimensions $n \times n$ and $n(n-1)/2 \times n(n-1)/2$, where $n = 2S+1$. Furthermore, the ordering of the elements of the matrix $\mathbf{M}^{(2)}$ is such that by reading across or down the final matrix, the successive lists of positions appear in lexicographic order. Now the traces $\text{Tr}(\mathbf{M}^{(1)})$ and $\text{Tr}(\mathbf{M}^{(2)})$ can be calculated analytically as

$$\text{Tr}(\mathbf{M}^{(1)}) = \frac{(-1)^{2S}}{\tau_N^{2S}} \sum_{i=0}^{2S}\left(\left(\prod_{s=1}^{i} p_s^-\right)\left(\prod_{r=i}^{2S-1} p_r^+\right)\right)$$

$$= \frac{(2S)!}{2^{2S}\tau_N^{2S}} \sum_{k=-S}^{S} e^{(k^2-S^2)\frac{\sigma}{S^2} + k\frac{\xi_0}{S}} = \frac{(2S)!e^{-\sigma}}{2^{2S}\tau_N^{2S}} Z_S$$

and

$$\mathrm{Tr}\left(\mathbf{M}^{(2)}\right) = \frac{(-1)^{2S+1}}{\tau_N^{2S-1}} \sum_{i=0}^{2S-1} \sum_{j=i+1}^{2S} \left( \prod_{s=1}^{i} p_s^- \prod_{r=j}^{2S-1} p_r^+ \sum_{m=1}^{j-i} \prod_{u=j+2-m}^{j} p_u^- \prod_{v=i}^{j-m-1} p_v^+ \right)$$

$$= \frac{(2S)! e^{-\sigma}}{2^{2S-1} \tau_N^{2S-1}} \sum_{k=-S}^{S-1} \sum_{n=k+1}^{S} \sum_{m=1}^{n-k} \frac{e^{[2k^2 - 2n - 1 + 2m(2n-m+1)]\frac{\sigma}{2S^2} + (2k+2m-1)\frac{\xi_0}{2S}}}{(S+n-m+1)(S-n+m)}.$$

Here we have used the result $\prod_{m=a}^{b} p_m^\pm = 1$ if $b < a$. Thus in the high barrier approximation, $\tau / \tau_N \approx \lambda_1^{-1}$ is given by the following approximate equation

$$\tau \approx \frac{2\tau_N}{Z_S} \sum_{k=-S}^{S-1} \sum_{n=k+1}^{S} \sum_{m=1}^{n-k} \frac{e^{[2k^2 - 2n - 1 + 2m(2n-m+1)]\frac{\sigma}{2S^2} + (2k+2m-1)\frac{\xi_0}{2S}}}{(S+n-m+1)(S-n+m)}. \tag{C6}$$